\documentclass[prd,twocolumn,floatfix,amsmath,nofootinbib,amssymb,floatfix]{revtex4}
\usepackage{graphicx,color,dcolumn,booktabs,bm}
\usepackage{longtable}
\usepackage{pdflscape}
\usepackage{pdfpages}
\usepackage{txfonts}
\usepackage{overpic}
\usepackage{amssymb}
\usepackage{makecell}
\usepackage{indentfirst}
\usepackage{feynmf}   
\usepackage{slashed}  
\usepackage{cases}
\usepackage{color}
\usepackage{multirow}
\usepackage{threeparttable}
\usepackage{epstopdf}
\usepackage{enumerate}
\usepackage{ulem}
\usepackage[figuresright]{rotating}
\usepackage[colorlinks,
            citecolor=blue,
            anchorcolor=red,
            menucolor=red,
            linkcolor=red,
            filecolor=red,
            runcolor=red,
            urlcolor=blue,
            frenchlinks=true]{hyperref}
\begin{document}

\title{Identifying triple-strangeness $\Omega$ hyperons in light of recent experimental results}
\author{Si-Qiang Luo$^{1,2,4}$}\email{luosq15@lzu.edu.cn}
\author{Xiang Liu$^{1,2,3,4}$}\email{xiangliu@lzu.edu.cn}
\affiliation{
$^1$School of Physical Science and Technology, Lanzhou University, Lanzhou 730000, China\\
$^2$Lanzhou Center for Theoretical Physics,
Key Laboratory of Theoretical Physics of Gansu Province,
Key Laboratory of Quantum Theory and Applications of MoE,
Gansu Provincial Research Center for Basic Disciplines of Quantum Physics, Lanzhou University, Lanzhou 730000, China\\
$^3$MoE Frontiers Science Center for Rare Isotopes, Lanzhou University, Lanzhou 730000, China\\
$^4$Research Center for Hadron and CSR Physics, Lanzhou University $\&$ Institute of Modern Physics of CAS, Lanzhou 730000, China}

\begin{abstract}
In this work, we systematically calculate the mass spectra and decay widths of $\Omega$ baryons with $N\leq3$, where $N=2n_\rho+2n_\lambda+l_\rho+l_\lambda$ is the shell of the three-body harmonic oscillator. In our scheme, the newly observed $\Omega(2109)$ is identified as a $N=2$ candidate, with its quantum numbers tentatively assigned as $\Omega(2S,3/2^+)$. However, the experimentally measured mass exhibits a deviation of approximately 50 MeV below the theoretical predictions. The possible reason for this deviation, along with the suggestion for experiments, are discussed in this paper. Additionally, we give possible assignments of $\Omega(2012)^-$, $\Omega(2250)^-$, $\Omega(2380)^-$, and $\Omega(2470)^-$. Furthermore, we calculate the masses and widths of other missing $\Omega$ states, which could provide valuable clues for experimental searches aimed at discovering these states.
\end{abstract}
\maketitle

\section{Introduction}\label{sec:Introduction}

In 1964, the establishment of the SU(3) symmetry for the classification of baryons was a milestone in the development of particle physics~\cite{Gell-Mann:1964ewy,Zweig:1964ruk,Zweig:1964jf}. A key experimental observation supporting this framework was the discovery of the $\Omega^-$ particle, as reported in Ref.~\cite{Barnes:1964pd}. Despite numerous hadrons being discovered in subsequent experiments, the number of identified $\Omega$ hyperons has remained limited, as reflected in the data collected by the Particle Data Group (PDG)~\cite{ParticleDataGroup:2024cfk}. During the past sixty years, only five $\Omega$ hyperons ($\Omega^-$, $\Omega(2012)^-$, $\Omega(2250)^-$, $\Omega(2380)^-$, and $\Omega(2470)^-$)~\cite{Barnes:1964pd,Aston:1987bb,Biagi:1985rn,Aston:1988yn,Belle:2018mqs} have been cataloged.

Recently, the BESIII Collaboration reported evidence for two excited $\Omega^-$ hyperons~\cite{BESIII:2024eqk}. The process $e^+e^- \to \Omega(2109)^- \bar{\Omega}^+ + c.c.$ revealed the $\Omega^*(2109)^-$ with a mass of $2108.5 \pm 5.2 \pm 0.9$ MeV and a width of $18.3 \pm 16.4 \pm 5.7$ MeV, observed with a significance of $4.1~\sigma$~\cite{BESIII:2024eqk}. Additionally, BESIII measured $e^+e^- \to \Omega(2012)^- \bar{\Omega}^+ + c.c.$ with 3.5 $\sigma$. The $\Omega(2012)^-$ was first identified by Belle in 2018~\cite{Belle:2018mqs}, which decays into $\Xi^0 K^-$ and $\Xi^- K_S^0$. These findings have spurred renewed interest in the spectroscopic behavior of $\Omega^-$ hyperons~\cite{Belle:2019zco,Belle:2022mrg,Xiao:2018pwe,Wang:2018hmi,Liu:2019wdr,Hyodo:2020czb,Hudspith:2024kzk,Hockley:2024aym,Crede:2024hur,Wang:2024ozz,Su:2024lzy,Aliev:2018syi,Valderrama:2018bmv,Lin:2018nqd,Huang:2018wth,Aliev:2018yjo,Zhong:2022cjx,Pavao:2018xub,Ikeno:2020vqv,Gutsche:2019eoh,Arifi:2022ntc,BESIII:2024eqk}.

\begin{figure}[htbp]
    \centering
    \includegraphics[width=8.6cm]{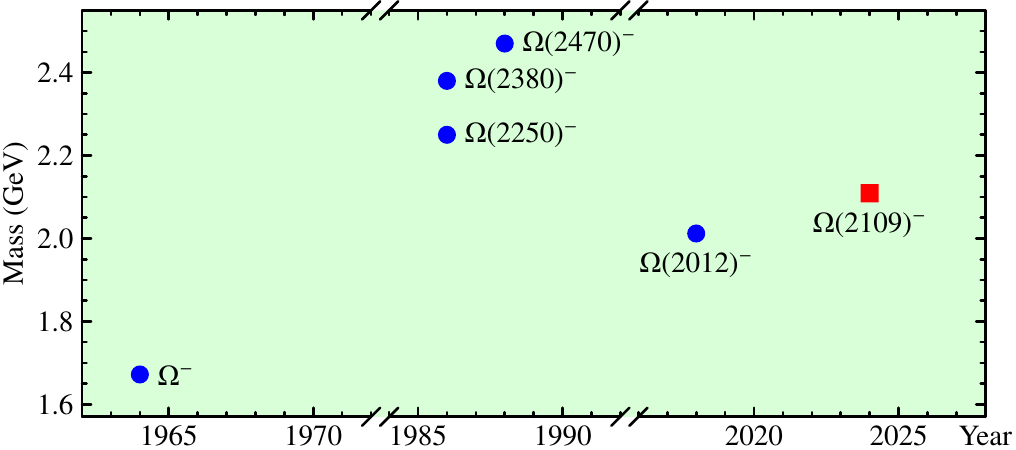}
    \caption{The observations of the $\Omega^-$ baryons~\cite{ParticleDataGroup:2024cfk,Barnes:1964pd,Belle:2018mqs, Aston:1987bb,Biagi:1985rn,Aston:1988yn,BESIII:2024eqk}.}
    \label{fig:observationsOmega}
\end{figure}

In Fig.~\ref{fig:observationsOmega}, the newly observed $\Omega(2109)^-$ is presented alongside the previously established $\Omega$ states, providing a comprehensive overview of the known $\Omega$ hyperons. Motivated by these recent experimental advances and recognizing that the $\Omega$ hyperon family remains incompletely established, we revisit the spectroscopic behavior of the $\Omega$ hyperons. With the advent of the high-precision era in hadron spectroscopy, we employ the Harmonic Oscillator Expansion Method to analyze the mass spectrum of $\Omega$ hyperons using a potential model~\cite{Luo:2019qkm}. This approach enables us to achieve higher precision in predicting the mass spectrum of systems with triply identical particles. Besides mass information being crucial for identifying $\Omega$ hyperons, we also calculate their two-body Okubo-Zweig-Iizuka (OZI)-allowed strong decays using the Quark Pair Creation (QPC) model~\cite{Micu:1968mk}. For these calculations, we use the spatial wave functions derived from the mass spectrum analysis as inputs.

Through these efforts, we aim to provide comprehensive guidance for the experimental search for excited $\Omega^-$ hyperons. In particular, we focus on identifying the properties of the newly reported $\Omega(2109)^-$~\cite{BESIII:2024eqk}, which we find can be interpreted as the $N=2$ candidate in the $\Omega^-$ hyperon family. This hypothesis can be further tested in future experiments.

This paper is organized as follows. After the Introduction, we present the mass spectrum of the $\Omega$ family in Sec.~\ref{sec:Mass_Spectrum}. Subsequently, we discuss the strong decay behaviors of the $\Omega$ states in Sec.~\ref{sec:Decay_Behavior}. This paper ends with a short summary in Sec.~\ref{sec:Summary}.

\section{Analysis of Mass Spectrum}\label{sec:Mass_Spectrum}

We analyze the mass spectrum using a non-relativistic potential model~\cite{Luo:2019qkm,Luo:2023sne}. This analysis is crucial not only for identifying the properties of the observed $\Omega$ states but also for providing essential input to study their decay properties. The Hamiltonian is given by
\begin{equation}\label{eq:H}
\hat{H}=\sum\limits_{i}T_i+\sum\limits_{i<j}V_{ij},
\end{equation}
where $T_i$ is defined as
\begin{equation}\label{eq:TiVij}
T_{i}=m_i+\frac{p_i^2}{2m_i}.
\end{equation}
In Eq.~(\ref{eq:TiVij}), $m_i$ and $p_i$ represent the mass and momentum of the constituent quark, respectively. The potential $V_{ij}$ in Eq.~(\ref{eq:H}) consists of the following components:
\begin{equation}
V_{ij}=V_{ij}^{\rm conf}+V_{ij}^{\rm hyp}+V_{ij}^{\rm so(cm)}+V_{ij}^{\rm so(tp)},
\end{equation}
which include the confinement potential
\begin{equation}
V_{ij}^{\rm conf}=-\frac{2}{3}\frac{\alpha_s}{r_{ij}}+\frac{b}{2}r_{ij}+\frac{1}{2}C,
\end{equation}
hyperfine interaction
\begin{equation}
V_{ij}^{\rm hyp}=\frac{2\alpha_s}{3m_im_j}\left[\frac{8\pi}{3}\tilde{\delta}(r_{ij}){\bf s}_i\cdot{\bf s}_j+\frac{1}{r_{ij}^3}S({\bf r},{\bf s}_i,{\bf s}_j)\right],
\end{equation}
\begin{equation}
\tilde{\delta}(r)=\frac{\sigma^3}{\pi^{3/2}}{\rm e}^{-\sigma^2r^2},~~~
S({\bf r},{\bf s}_i,{\bf s}_j)=\frac{3{\bf s}_i\cdot{\bf r}_{ij}{\bf s}_j\cdot{\bf r}_{ij}}{r_{ij}^2}-{\bf s}_i\cdot{\bf s}_j,
\end{equation}
color-magnetic spin-orbit term
\begin{equation}
\begin{split}
V_{ij}^{{\rm so(cm)}}=&\frac{2\alpha_s}{3r_{ij}^3}\left(\frac{{\bf r}_{ij}\times{\bf p}_i\cdot{\bf s}_i}{m_i^2}-\frac{{\bf r}_{ij}\times{\bf p}_j\cdot{\bf s}_j}{m_j^2}\right.\\
&\left.-\frac{{\bf r}_{ij}\times{\bf p}_j\cdot{\bf s}_i-{\bf r}_{ij}\times{\bf p}_i\cdot{\bf s}_j}{m_im_j}\right),
\end{split}
\end{equation}
and Thomas-precession spin-orbit term
\begin{equation}
V_{ij}^{{\rm so(tp)}}=-\frac{1}{2r_{ij}}\frac{\partial H_{ij}^{\rm conf}}{\partial r_{ij}}\left(\frac{{\bf r}_{ij}\times{\bf p}_i\cdot{\bf s}_i}{m_i^2}-\frac{{\bf r}_{ij}\times{\bf p}_j\cdot{\bf s}_j}{m_j^2}\right).
\end{equation}
In the above potentials, $\alpha_s$, $b$, $C$, and $\sigma$ denote the coupling constant of the one-gluon exchange, confinement strength, mass-normalization constant, and smearing parameter, respectively.

Since an $\Omega$ baryon consists of three identical strange constituent quarks, its wave function must be  fully anti-symmetric. In general, the baryon wave functions comprises color, flavor, spin, and orbital parts. For a given $J^P$, the sub-basis can be expressed as
\begin{equation}\label{eq:phialpha}
|\phi^\alpha_{J^PM_J}\rangle=|\phi^{\rm color}\phi^{\rm flavor}[[R_{n_\rho l_\rho}^\rho R_{n_\lambda l_\lambda}^\lambda]_{L}\chi_{s_{12}S}]_{JM_J}\rangle,
\end{equation}
where
\begin{equation}
\phi^{\rm color}=\frac{rgb-rbg+gbr-grb+brg-bgr}{\sqrt{6}}
\end{equation}
is color wave function,
\begin{equation}
\phi^{\rm flavor}=sss    
\end{equation}
is flavor wave function,
\begin{equation}
|\chi_{s_{12}S}\rangle=|[[s_1s_2]_{s_{12}}s_3]_S\rangle
\end{equation}
is the spin wave function, and $|[R_{n_\rho l_\rho}^\rho R_{n_\lambda l_\lambda}^\lambda]_{L}\rangle$ is orbital wave function. For the orbital wave function, we use $\rho$- and $\lambda$-modes to describe the three-body system, with coordinates defined as
\begin{equation}
\begin{split}
{\bm \rho}   =&{\bf r}_1-{\bf r}_2,\\
{\bm \lambda}=&\frac{m_1{\bf r}_1+m_2{\bf r}_2}{m_1+m_2}-{\bf r}_3.
\end{split}
\end{equation}
For the $\Omega$ baryon, $m_1=m_2=m_3$, simplifying the coordinates to
\begin{equation}
\begin{split}
{\bm \rho}   =&{\bf r}_1-{\bf r}_2,\\
{\bm \lambda}=&\frac{{\bf r}_1+{\bf r}_2}{2}-{\bf r}_3.
\end{split}
\end{equation}
In this work, we employ harmonic oscillator wave function. In coordinate space, harmonic oscillator wave function is 
\begin{equation}\label{eq:shor}
\begin{split}
R_{nlm}(\beta,{\bf r})=&\beta^{l+\frac{3}{2}}\sqrt{\frac{2n!}{\Gamma(n+l+\frac{3}{2})}}L_{n}^{l+\frac{1}{2}}(\beta^2r^2)\\
&\times r^l{\rm e}^{-\frac{\beta^2 r^2}{2}}Y_{lm}(\hat{\bf r}),
\end{split}
\end{equation}
where $n$, $l$ are radial and orbital quantum numbers, respectively and $m$ is the third component of the orbital angular momentum. In momentum space, the expression is
\begin{equation}\label{eq:shop}
\begin{split}
R_{nlm}(\beta,{\bf p})=&\frac{(-1)^n(-{\mathrm i})^l}{\beta^{\frac{3}{2}+l}}\sqrt{\frac{2n!}{\Gamma(n+l+\frac{3}{2})}}L_{n}^{l+\frac{1}{2}}({p^2}/{\beta^2})\\
&\times p^l{\rm e}^{-\frac{p^2}{2\beta^2}} Y_{l m}(\hat{\bf p}),
\end{split}
\end{equation}
The orbital wave function $|[R_{n_\rho l_\rho}^\rho R_{n_\lambda l_\lambda}^\lambda]_{LM_L}\rangle$ is defined as
\begin{equation}
|[R_{n_\rho l_\rho}^\rho R_{n_\lambda l_\lambda}^\lambda]_{LM_L}\rangle=|[R_{n_\rho l_\rho}(\beta_\rho,{\bm \rho})R_{n_\lambda l_\lambda}(\beta_\lambda,{\bm \lambda})]_{LM_L}\rangle.
\end{equation}
Following the above conventions, we use $\alpha=\{n_\rho l_\rho, n_\lambda l_\lambda, L,s_{12},S\}$ to denote the list of quantum numbers.

\begin{table*}[htbp]
\caption{The spin-spatial parts of the basis states for the $\Omega$ baryons with $ N \leq 3 $ are presented below. Since all states share the same color and flavor wave functions, $\phi^{\rm color}$ and $\phi^{\rm flavor}$ are omitted for simplicity. The subscripts ``1'' and ``2'' indicate that the same $|\psi_{J^P}^{NLS}\rangle$ may contain multiple basis states.}
\label{tab:base}
\renewcommand\arraystretch{1.75}
\begin{tabular*}{178mm}{@{\extracolsep{\fill}}ccll}
\toprule[1.00pt]
\toprule[1.00pt]
$N$ &$J^P$           &\multicolumn{1}{c}{$|\psi^{NLS}_{J^P}\rangle$} &\multicolumn{1}{c}{Basis}                                                                                                                                                                                                                                                                                          \\
\midrule[0.75pt]
0   &$\frac{3}{2}^+$ &$|\psi^{0,0,3/2}_{3/2^+}\rangle$               &$|[[R_{00}^\rho R_{00}^\lambda]_0\chi_{1,\frac{3}{2}}]_\frac{3}{2}\rangle$                                                                                                                                                                                                                                        \\
1   &$\frac{1}{2}^-$ &$|\psi^{1,1,1/2}_{1/2^-}\rangle$               &$\frac{1}{\sqrt{2}}|[[R_{00}^\rho R_{01}^\lambda]_1\chi_{1,\frac{1}{2}}]_\frac{1}{2}\rangle+\frac{1}{\sqrt{2}}|[[R_{01}^\rho R_{00}^\lambda]_1\chi_{0,\frac{1}{2}}]_\frac{1}{2}\rangle$                                                                                                                           \\
1   &$\frac{3}{2}^-$ &$|\psi^{1,1,1/2}_{3/2^-}\rangle$               &$\frac{1}{\sqrt{2}}|[[R_{00}^\rho R_{01}^\lambda]_1\chi_{1,\frac{1}{2}}]_\frac{3}{2}\rangle+\frac{1}{\sqrt{2}}|[[R_{01}^\rho R_{00}^\lambda]_1\chi_{0,\frac{1}{2}}]_\frac{3}{2}\rangle$                                                                                                                           \\
2   &$\frac{1}{2}^+$ &$|\psi^{2,0,1/2}_{1/2^+}\rangle$               &$\frac{1}{\sqrt{2}}|[[R_{01}^\rho R_{01}^\lambda]_0\chi_{0,\frac{1}{2}}]_\frac{1}{2}\rangle-\frac{1}{2}|[[R_{00}^\rho R_{10}^\lambda]_0\chi_{1,\frac{1}{2}}]_\frac{1}{2}\rangle+\frac{1}{2}|[[R_{10}^\rho R_{00}^\lambda]_0\chi_{1,\frac{1}{2}}]_\frac{1}{2}\rangle$                                              \\
2   &$\frac{1}{2}^+$ &$|\psi^{2,2,3/2}_{1/2^+}\rangle$               &$\frac{1}{\sqrt{2}}|[[R_{00}^\rho R_{02}^\lambda]_2\chi_{1,\frac{3}{2}}]_\frac{1}{2}\rangle+\frac{1}{\sqrt{2}}|[[R_{02}^\rho R_{00}^\lambda]_2\chi_{1,\frac{3}{2}}]_\frac{1}{2}\rangle$                                                                                                                           \\
2   &$\frac{3}{2}^+$ &$|\psi^{2,0,3/2}_{3/2^+}\rangle$               &$\frac{1}{\sqrt{2}}|[[R_{00}^\rho R_{10}^\lambda]_0\chi_{1,\frac{3}{2}}]_\frac{3}{2}\rangle+\frac{1}{\sqrt{2}}|[[R_{10}^\rho R_{00}^\lambda]_0\chi_{1,\frac{3}{2}}]_\frac{3}{2}\rangle$                                                                                                                           \\
2   &$\frac{3}{2}^+$ &$|\psi^{2,2,1/2}_{3/2^+}\rangle$               &$\frac{1}{2}|[[R_{00}^\rho R_{02}^\lambda]_2\chi_{1,\frac{1}{2}}]_\frac{3}{2}\rangle-\frac{1}{\sqrt{2}}|[[R_{01}^\rho R_{01}^\lambda]_2\chi_{0,\frac{1}{2}}]_\frac{3}{2}\rangle-\frac{1}{2}|[[R_{02}^\rho R_{00}^\lambda]_2\chi_{1,\frac{1}{2}}]_\frac{3}{2}\rangle$                                              \\
2   &$\frac{3}{2}^+$ &$|\psi^{2,2,3/2}_{3/2^+}\rangle$               &$\frac{1}{\sqrt{2}}|[[R_{00}^\rho R_{02}^\lambda]_2\chi_{1,\frac{3}{2}}]_\frac{3}{2}\rangle+\frac{1}{\sqrt{2}}|[[R_{02}^\rho R_{00}^\lambda]_2\chi_{1,\frac{3}{2}}]_\frac{3}{2}\rangle$                                                                                                                           \\
2   &$\frac{5}{2}^+$ &$|\psi^{2,2,1/2}_{5/2^+}\rangle$               &$\frac{1}{2}|[[R_{00}^\rho R_{02}^\lambda]_2\chi_{1,\frac{1}{2}}]_\frac{5}{2}\rangle-\frac{1}{\sqrt{2}}|[[R_{01}^\rho R_{01}^\lambda]_2\chi_{0,\frac{1}{2}}]_\frac{5}{2}\rangle-\frac{1}{2}|[[R_{02}^\rho R_{00}^\lambda]_2\chi_{1,\frac{1}{2}}]_\frac{5}{2}\rangle$                                              \\
2   &$\frac{5}{2}^+$ &$|\psi^{2,2,3/2}_{5/2^+}\rangle$               &$\frac{1}{\sqrt{2}}|[[R_{00}^\rho R_{02}^\lambda]_2\chi_{1,\frac{3}{2}}]_\frac{5}{2}\rangle+\frac{1}{\sqrt{2}}|[[R_{02}^\rho R_{00}^\lambda]_2\chi_{1,\frac{3}{2}}]_\frac{5}{2}\rangle$                                                                                                                           \\
2   &$\frac{7}{2}^+$ &$|\psi^{2,2,3/2}_{7/2^+}\rangle$               &$\frac{1}{\sqrt{2}}|[[R_{00}^\rho R_{02}^\lambda]_2\chi_{1,\frac{3}{2}}]_\frac{7}{2}\rangle+\frac{1}{\sqrt{2}}|[[R_{02}^\rho R_{00}^\lambda]_2\chi_{1,\frac{3}{2}}]_\frac{7}{2}\rangle$                                                                                                                           \\
3   &$\frac{1}{2}^-$ &$|\psi^{3,1,1/2}_{1/2^-}\rangle_{1}$           &$-\frac{\sqrt{\frac{5}{2}}}{3}|[[R_{01}^\rho R_{02}^\lambda]_1\chi_{0,\frac{1}{2}}]_\frac{1}{2}\rangle-\frac{\sqrt{\frac{5}{2}}}{3}|[[R_{02}^\rho R_{01}^\lambda]_1\chi_{1,\frac{1}{2}}]_\frac{1}{2}\rangle+\frac{\sqrt{2}}{3}|[[R_{01}^\rho R_{10}^\lambda]_1\chi_{0,\frac{1}{2}}]_\frac{1}{2}\rangle$           \\
    &                &                                               &$+\frac{\sqrt{2}}{3}|[[R_{10}^\rho R_{01}^\lambda]_1\chi_{1,\frac{1}{2}}]_\frac{1}{2}\rangle$                                                                                                                                                                                                                     \\
\addlinespace[0.55em]
3   &$\frac{1}{2}^-$ &$|\psi^{3,1,1/2}_{1/2^-}\rangle_{2}$           &$-\frac{1}{3 \sqrt{2}}|[[R_{01}^\rho R_{02}^\lambda]_1\chi_{0,\frac{1}{2}}]_\frac{1}{2}\rangle-\frac{1}{3 \sqrt{2}}|[[R_{02}^\rho R_{01}^\lambda]_1\chi_{1,\frac{1}{2}}]_\frac{1}{2}\rangle-\frac{\sqrt{\frac{3}{2}}}{2}|[[R_{00}^\rho R_{11}^\lambda]_1\chi_{1,\frac{1}{2}}]_\frac{1}{2}\rangle$                 \\
    &                &                                               &$-\frac{\sqrt{\frac{5}{2}}}{6}|[[R_{01}^\rho R_{10}^\lambda]_1\chi_{0,\frac{1}{2}}]_\frac{1}{2}\rangle-\frac{\sqrt{\frac{5}{2}}}{6}|[[R_{10}^\rho R_{01}^\lambda]_1\chi_{1,\frac{1}{2}}]_\frac{1}{2}\rangle-\frac{\sqrt{\frac{3}{2}}}{2}|[[R_{11}^\rho R_{00}^\lambda]_1\chi_{0,\frac{1}{2}}]_\frac{1}{2}\rangle$ \\
\addlinespace[0.55em]
3   &$\frac{1}{2}^-$ &$|\psi^{3,1,3/2}_{1/2^-}\rangle$               &$\frac{1}{\sqrt{3}}|[[R_{02}^\rho R_{01}^\lambda]_1\chi_{1,\frac{3}{2}}]_\frac{1}{2}\rangle-\frac{1}{2}|[[R_{00}^\rho R_{11}^\lambda]_1\chi_{1,\frac{3}{2}}]_\frac{1}{2}\rangle+\frac{\sqrt{\frac{5}{3}}}{2}|[[R_{10}^\rho R_{01}^\lambda]_1\chi_{1,\frac{3}{2}}]_\frac{1}{2}\rangle$                             \\
3   &$\frac{3}{2}^-$ &$|\psi^{3,1,1/2}_{3/2^-}\rangle_{1}$           &$-\frac{\sqrt{\frac{5}{2}}}{3}|[[R_{01}^\rho R_{02}^\lambda]_1\chi_{0,\frac{1}{2}}]_\frac{3}{2}\rangle-\frac{\sqrt{\frac{5}{2}}}{3}|[[R_{02}^\rho R_{01}^\lambda]_1\chi_{1,\frac{1}{2}}]_\frac{3}{2}\rangle+\frac{\sqrt{2}}{3}|[[R_{01}^\rho R_{10}^\lambda]_1\chi_{0,\frac{1}{2}}]_\frac{3}{2}\rangle$           \\
    &                &                                               &$+\frac{\sqrt{2}}{3}|[[R_{10}^\rho R_{01}^\lambda]_1\chi_{1,\frac{1}{2}}]_\frac{3}{2}\rangle$                                                                                                                                                                                                                     \\
\addlinespace[0.55em]
3   &$\frac{3}{2}^-$ &$|\psi^{3,1,1/2}_{3/2^-}\rangle_{2}$           &$-\frac{1}{3 \sqrt{2}}|[[R_{01}^\rho R_{02}^\lambda]_1\chi_{0,\frac{1}{2}}]_\frac{3}{2}\rangle-\frac{1}{3 \sqrt{2}}|[[R_{02}^\rho R_{01}^\lambda]_1\chi_{1,\frac{1}{2}}]_\frac{3}{2}\rangle-\frac{\sqrt{\frac{3}{2}}}{2}|[[R_{00}^\rho R_{11}^\lambda]_1\chi_{1,\frac{1}{2}}]_\frac{3}{2}\rangle$                 \\
    &                &                                               &$-\frac{\sqrt{\frac{5}{2}}}{6}|[[R_{01}^\rho R_{10}^\lambda]_1\chi_{0,\frac{1}{2}}]_\frac{3}{2}\rangle-\frac{\sqrt{\frac{5}{2}}}{6}|[[R_{10}^\rho R_{01}^\lambda]_1\chi_{1,\frac{1}{2}}]_\frac{3}{2}\rangle-\frac{\sqrt{\frac{3}{2}}}{2}|[[R_{11}^\rho R_{00}^\lambda]_1\chi_{0,\frac{1}{2}}]_\frac{3}{2}\rangle$ \\
\addlinespace[0.55em]
3   &$\frac{3}{2}^-$ &$|\psi^{3,1,3/2}_{3/2^-}\rangle$               &$\frac{1}{\sqrt{3}}|[[R_{02}^\rho R_{01}^\lambda]_1\chi_{1,\frac{3}{2}}]_\frac{3}{2}\rangle-\frac{1}{2}|[[R_{00}^\rho R_{11}^\lambda]_1\chi_{1,\frac{3}{2}}]_\frac{3}{2}\rangle+\frac{\sqrt{\frac{5}{3}}}{2}|[[R_{10}^\rho R_{01}^\lambda]_1\chi_{1,\frac{3}{2}}]_\frac{3}{2}\rangle$                             \\
3   &$\frac{3}{2}^-$ &$|\psi^{3,2,1/2}_{3/2^-}\rangle$               &$-\frac{1}{\sqrt{2}}|[[R_{01}^\rho R_{02}^\lambda]_2\chi_{0,\frac{1}{2}}]_\frac{3}{2}\rangle+\frac{1}{\sqrt{2}}|[[R_{02}^\rho R_{01}^\lambda]_2\chi_{1,\frac{1}{2}}]_\frac{3}{2}\rangle$                                                                                                                          \\
3   &$\frac{3}{2}^-$ &$|\psi^{3,3,3/2}_{3/2^-}\rangle$               &$-\frac{1}{2}|[[R_{00}^\rho R_{03}^\lambda]_3\chi_{1,\frac{3}{2}}]_\frac{3}{2}\rangle+\frac{\sqrt{3}}{2}|[[R_{02}^\rho R_{01}^\lambda]_3\chi_{1,\frac{3}{2}}]_\frac{3}{2}\rangle$                                                                                                                                 \\
3   &$\frac{5}{2}^-$ &$|\psi^{3,1,3/2}_{5/2^-}\rangle$               &$\frac{1}{\sqrt{3}}|[[R_{02}^\rho R_{01}^\lambda]_1\chi_{1,\frac{3}{2}}]_\frac{5}{2}\rangle-\frac{1}{2}|[[R_{00}^\rho R_{11}^\lambda]_1\chi_{1,\frac{3}{2}}]_\frac{5}{2}\rangle+\frac{\sqrt{\frac{5}{3}}}{2}|[[R_{10}^\rho R_{01}^\lambda]_1\chi_{1,\frac{3}{2}}]_\frac{5}{2}\rangle$                             \\
3   &$\frac{5}{2}^-$ &$|\psi^{3,2,1/2}_{5/2^-}\rangle$               &$-\frac{1}{\sqrt{2}}|[[R_{01}^\rho R_{02}^\lambda]_2\chi_{0,\frac{1}{2}}]_\frac{5}{2}\rangle+\frac{1}{\sqrt{2}}|[[R_{02}^\rho R_{01}^\lambda]_2\chi_{1,\frac{1}{2}}]_\frac{5}{2}\rangle$                                                                                                                          \\
3   &$\frac{5}{2}^-$ &$|\psi^{3,3,1/2}_{5/2^-}\rangle$               &$-\frac{\sqrt{\frac{3}{2}}}{2}|[[R_{00}^\rho R_{03}^\lambda]_3\chi_{1,\frac{1}{2}}]_\frac{5}{2}\rangle-\frac{1}{2 \sqrt{2}}|[[R_{01}^\rho R_{02}^\lambda]_3\chi_{0,\frac{1}{2}}]_\frac{5}{2}\rangle-\frac{1}{2 \sqrt{2}}|[[R_{02}^\rho R_{01}^\lambda]_3\chi_{1,\frac{1}{2}}]_\frac{5}{2}\rangle$                 \\
    &                &                                               &$-\frac{\sqrt{\frac{3}{2}}}{2}|[[R_{03}^\rho R_{00}^\lambda]_3\chi_{0,\frac{1}{2}}]_\frac{5}{2}\rangle$                                                                                                                                                                                                           \\
\addlinespace[0.55em]
3   &$\frac{5}{2}^-$ &$|\psi^{3,3,3/2}_{5/2^-}\rangle$               &$-\frac{1}{2}|[[R_{00}^\rho R_{03}^\lambda]_3\chi_{1,\frac{3}{2}}]_\frac{5}{2}\rangle+\frac{\sqrt{3}}{2}|[[R_{02}^\rho R_{01}^\lambda]_3\chi_{1,\frac{3}{2}}]_\frac{5}{2}\rangle$                                                                                                                                 \\
3   &$\frac{7}{2}^-$ &$|\psi^{3,3,1/2}_{7/2^-}\rangle$               &$-\frac{\sqrt{\frac{3}{2}}}{2}|[[R_{00}^\rho R_{03}^\lambda]_3\chi_{1,\frac{1}{2}}]_\frac{7}{2}\rangle-\frac{1}{2 \sqrt{2}}|[[R_{01}^\rho R_{02}^\lambda]_3\chi_{0,\frac{1}{2}}]_\frac{7}{2}\rangle-\frac{1}{2 \sqrt{2}}|[[R_{02}^\rho R_{01}^\lambda]_3\chi_{1,\frac{1}{2}}]_\frac{7}{2}\rangle$                 \\
    &                &                                               &$-\frac{\sqrt{\frac{3}{2}}}{2}|[[R_{03}^\rho R_{00}^\lambda]_3\chi_{0,\frac{1}{2}}]_\frac{7}{2}\rangle$                                                                                                                                                                                                           \\
\addlinespace[0.55em]
3   &$\frac{7}{2}^-$ &$|\psi^{3,3,3/2}_{7/2^-}\rangle$               &$-\frac{1}{2}|[[R_{00}^\rho R_{03}^\lambda]_3\chi_{1,\frac{3}{2}}]_\frac{7}{2}\rangle+\frac{\sqrt{3}}{2}|[[R_{02}^\rho R_{01}^\lambda]_3\chi_{1,\frac{3}{2}}]_\frac{7}{2}\rangle$                                                                                                                                 \\
3   &$\frac{9}{2}^-$ &$|\psi^{3,3,3/2}_{9/2^-}\rangle$               &$-\frac{1}{2}|[[R_{00}^\rho R_{03}^\lambda]_3\chi_{1,\frac{3}{2}}]_\frac{9}{2}\rangle+\frac{\sqrt{3}}{2}|[[R_{02}^\rho R_{01}^\lambda]_3\chi_{1,\frac{3}{2}}]_\frac{9}{2}\rangle$                                                                                                                                 \\
\bottomrule[1.00pt]
\bottomrule[1.00pt]
\end{tabular*}
\end{table*}

It is obvious that we could not ensure the wave function is fully anti-symmetric with only one basis with Eq.~(\ref{eq:phialpha}). We consider a linear combination of Eq.~(\ref{eq:phialpha}) with a series of $\alpha$ and the corresponding coefficients. We first impose the condition $(-1)^{l_\rho +s_{12}}=-1$, which ensures the anti-symmetry of the two-quark subsystem at $\rho$ mode. We then introduce the operator $\hat{\cal P}_{13}$ to act on the basis state $|\phi^\alpha_{J^P M}\rangle$, where $\hat{\cal P}_{13}$ could exchange the wave functions of quark-1 and quark-3. Since the wave functions of quark-1 and quark-2 are automatically anti-symmetric, if we construct some special expressions to make the wave functions of quark-1 and quark-3 also anti-symmetric, we could ensure the fully anti-symmetric wave functions. After performing the $\hat{\cal P}_{13}$ with $|\phi^\alpha_{J^P M}\rangle$, the following properties of $\hat{\cal P}_{13}$ are obtained
\begin{enumerate}
\item The operator $\hat{\cal P}_{13}$ does not change the energy level $N=2n_\rho+l_\rho+2n_\lambda+l_\lambda$. 
\item The total quantum number $L$ remains unchanged under $\hat{\cal P}_{13}$.
\item The total spin $S$ is also invariant under $\hat{\cal P}_{13}$.
\end{enumerate}
In short summary, the $\hat{\cal P}_{13}$ could not change the quantum numbers $N$, $L$, and $S$. Based on this point, we use the $NLS$ basis to describe the $\Omega$ baryons:
\begin{equation}\label{eq:psiNLS}
|\psi^{NLS}_{J^PM}\rangle=\sum\limits_{i} c_i|\phi^{\alpha_i}_{J^PM}\rangle,
\end{equation}
which may contains a series of sub basis $|\phi^{\alpha_i}_{J^PM}\rangle$. The quantum numbers $\{n_\rho l_\rho, n_\lambda l_\lambda, L,s_{12},S\}$ in $\alpha^i$ should satisfy
\begin{equation}
\makecell[c]{
(-1)^{l_\rho+l_\lambda}=P,\\
2n_\rho+l_\rho+2n_\lambda+l_\lambda=N,\\
|l_\rho-l_\lambda|\leq L \leq |l_\rho+l_\lambda|,\\
|s_{12}-1/2|\leq S \leq |s_{12}+1/2|,\\
|L-S|\leq J \leq |L+S|.
}
\end{equation}

We define the vector
\begin{equation}\label{eq:vector}
v=\left(\begin{array}{cccc}
|\phi^{\alpha_1}_{J^PM}\rangle & |\phi^{\alpha_2}_{J^PM}\rangle & |\phi^{\alpha_3}_{J^PM}\rangle & \cdots
\end{array}\right)^T.
\end{equation}
The matrix $\cal P$ is then constructed as
\begin{equation}
{\cal P}=v \hat{\cal P}_{13} v^T.
\end{equation}
If the eigenvalues of $\cal P$ are $-1$, the corresponding eigenvector $c_i$ are used in Eq.~(\ref{eq:psiNLS}) to ensure
\begin{equation}
\hat{\cal P}_{13}|\psi^{NLS}_{JM}\rangle=-|\psi^{NLS}_{JM}\rangle.
\end{equation}

The action of $\hat{\cal P}_{13}$ on the wave function components is as follows:
\begin{enumerate}
\item For the color wave function:
\begin{equation}\label{eq:P13color}
\langle \phi^{\rm color}|\hat{\cal P}_{13}|\phi^{\rm color}\rangle=-1.
\end{equation}
\item For the flavor wave function:
\begin{equation}
\langle \phi^{\rm flavor}|\hat{\cal P}_{13}|\phi^{\rm flavor}\rangle=1.
\end{equation}
\item For the spin wave function:
\begin{equation}
\langle \chi_{1\frac{3}{2}}|\hat{\cal P}_{13}|\chi_{1\frac{3}{2}}\rangle=1,
\end{equation}
and
\begin{equation}
\left(
\begin{array}{c}
\langle \chi_{0\frac{1}{2}}|\\
\langle \chi_{1\frac{1}{2}}|
\end{array}
\right)\hat{\cal P}_{13}\left(\begin{array}{cc}
| \chi_{0\frac{1}{2}}\rangle&
|\chi_{1\frac{1}{2}}\rangle
\end{array}\right)=\left(
\begin{array}{cc}
\frac{1}{2}     &-\frac{\sqrt{3}}{2}  \\
-\frac{\sqrt{3}}{2}     &-\frac{1}{2} 
\end{array}
\right).
\end{equation}
\item

For the orbital wave function, the $\beta$ values must satisfy $\beta_\rho =\frac{2}{\sqrt{3}}\beta_\lambda$ according to the solutions of three-body Schr\"odinger equation of harmonic oscillator with three identical particles\footnote{In some works, the Jacobi coordinates are defined as ${\bm \rho}=\frac{1}{\sqrt{2}}({\bf r}_1-{\bf r}_2)$ and ${\bm \lambda}=\sqrt{\frac{2}{3}}(\frac{{\bf r}_1+{\bf r}_2}{2}-{\bf r}_3)$~\cite{Wang:2018hmi,Liu:2019wdr,Zhong:2024mnt,Wang:2024ozz,Richard:1992uk,Capstick:1986ter,Murthy:1985fd}. The corresponding transformations are $\hat{\cal P}_{13}{\bm \rho}=\frac{1}{2}{\bm \rho}-\frac{\sqrt{3}}{2}{\bm \lambda}$ and $\hat{\cal P}_{13}{\bm \lambda}=-\frac{\sqrt{3}}{2}{\bm \rho}-\frac{1}{2}{\bm \lambda}$~\cite{Richard:1992uk}. With this definition, the $\beta$ values satisfy $\beta_\rho=\beta_\lambda$.}. This leads to
\begin{equation}
\begin{split}
\hat{\cal P}_{13}{\bm \rho   }=&\frac{1}{2}{\bm \rho}-{\bm \lambda},\\
\hat{\cal P}_{13}{\bm \lambda}=&-\frac{3}{4}{\bm \rho}-\frac{1}{2}{\bm \lambda}.\\
\end{split}
\end{equation}
Similar to Ref.~\cite{Richard:1992uk}, the matrix elements of $\hat{\cal P}_{13}$ for the orbital wave functions are
\begin{equation}\label{eq:P13spatial}
\begin{split}
&\langle [R_{n_\rho^\prime l_\rho^\prime}^\rho R_{n_\lambda^\prime l_\lambda^\prime}^\lambda]_{LM}|\hat{\cal P}_{13}|[R_{n_\rho l_\rho}^\rho R_{n_\lambda l_\lambda}^\lambda]_{LM}\rangle\\
=&\int{\rm d}^3{\bm \rho}{\rm d}^3{\bm \lambda} [R_{n_\rho^\prime l_\rho^\prime}(\beta_\rho,{\bm \rho})R_{n_\lambda^\prime l_\lambda^\prime}(\beta_\lambda,{\bm \lambda})]^*_{LM}\\
&\times [R_{n_\rho l_\rho}(\beta_\rho,\frac{1}{2}{\bm \rho}-{\bm \lambda})R_{n_\lambda l_\lambda}(\beta_\lambda,-\frac{3}{4}{\bm \rho}-\frac{1}{2}{\bm \lambda})]_{LM}.
\end{split}
\end{equation}
\end{enumerate}

In calculations, we chose $N\leq 9$ for $P=-1$ and $N\leq 10$ for $P=+1$ to ensure a complete set of solutions as possible. For example, for $J^P=\frac{3}{2}^+$ states, the finial basis states are
\begin{equation}
\psi=\left(\begin{array}{ccccc}
|\psi_{3/2^+}^{0,0,3/2}\rangle&
|\psi_{3/2^+}^{2,0,3/2}\rangle&
|\psi_{3/2^+}^{2,2,1/2}\rangle&
|\psi_{3/2^+}^{2,2,3/2}\rangle&
\cdots
\end{array}\right)^T.
\end{equation}
For $|\psi_{3/2^+}^{0,0,3/2}\rangle$, we find only one vector defined in Eq.~(\ref{eq:vector}), i.e.,
\begin{equation}
v=\left(|\phi_{3/2^+M}^{\alpha_1}\rangle\right)=\left(|\phi^{\rm color}\phi^{\rm flavor}[[R_{00}^\rho R_{00}^\lambda]_0\chi_{1,\frac{3}{2}}]_\frac{3}{2}\rangle\right).
\end{equation}
With Eqs.~(\ref{eq:P13color})-(\ref{eq:P13spatial}), one could obtained the matrix ${\cal P}$ with
\begin{equation}
{\cal P}=\left(-1\right).
\end{equation}
The only eigenvalue is $-1$, and the corresponding eigenvector is $c=(1)$. If we take the $c$ to Eq.~(\ref{eq:psiNLS}), one could obtain
\begin{equation}
|\psi_{3/2^+}^{0,0,3/2}\rangle=|\phi^{\rm color}\phi^{\rm flavor}[[R_{00}^\rho R_{00}^\lambda]_0\chi_{1,\frac{3}{2}}]_\frac{3}{2}\rangle.
\end{equation}
For the $|\psi_{3/2^+}^{2,0,3/2}\rangle$, the vector is
\begin{equation}
v=\left(\begin{array}{c}
|\phi^{\rm color}\phi^{\rm flavor}[[R_{00}^\rho R_{10}^\lambda]_0\chi_{1,\frac{3}{2}}]_\frac{3}{2}\rangle\\
|\phi^{\rm color}\phi^{\rm flavor}[[R_{10}^\rho R_{00}^\lambda]_0\chi_{1,\frac{3}{2}}]_\frac{3}{2}\rangle
\end{array}\right),
\end{equation}
and the matrix ${\cal P}$ is
\begin{equation}
{\cal P}=\left(
\begin{array}{cc}
 -\frac{1}{4} & -\frac{3}{4} \\
 -\frac{3}{4} & -\frac{1}{4} \\
\end{array}
\right).
\end{equation}
The eigenvector corresponding to $-1$ is $c=(1/\sqrt{2},1/\sqrt{2})$. In addition, for the $|\psi_{3/2^+}^{2,2,1/2}\rangle$, the vector, matrix ${\cal P}$, and eigenvector are
\begin{equation}
v=\left(\begin{array}{c}
|\phi^{\rm color}\phi^{\rm flavor}[[R_{00}^\rho R_{02}^\lambda]_2\chi_{1,\frac{1}{2}}]_\frac{3}{2}\rangle\\
|\phi^{\rm color}\phi^{\rm flavor}[[R_{01}^\rho R_{01}^\lambda]_2\chi_{0,\frac{1}{2}}]_\frac{3}{2}\rangle\\
|\phi^{\rm color}\phi^{\rm flavor}[[R_{02}^\rho R_{00}^\lambda]_2\chi_{1,\frac{1}{2}}]_\frac{3}{2}\rangle
\end{array}\right),
\end{equation}
\begin{equation}
{\cal P}=\left(
\begin{array}{ccc}
 \frac{1}{8} & \frac{3}{4 \sqrt{2}} & \frac{3}{8} \\
 \frac{3}{4 \sqrt{2}} & -\frac{1}{4} & -\frac{3}{4 \sqrt{2}} \\
 \frac{3}{8} & -\frac{3}{4 \sqrt{2}} & \frac{1}{8} \\
\end{array}
\right),
\end{equation}
and
\begin{equation}
c=(1/2,-1/\sqrt{2},-1/2),
\end{equation}
respectively. For the $|\psi_{3/2^+}^{2,2,3/2}\rangle$, we also have
\begin{equation}
v=\left(\begin{array}{c}
|\phi^{\rm color}\phi^{\rm flavor}[[R_{00}^\rho R_{02}^\lambda]_2\chi_{1,\frac{3}{2}}]_\frac{3}{2}\rangle\\
|\phi^{\rm color}\phi^{\rm flavor}[[R_{02}^\rho R_{00}^\lambda]_2\chi_{1,\frac{3}{2}}]_\frac{3}{2}\rangle
\end{array}\right),
\end{equation}
\begin{equation}
{\cal P}=\left(
\begin{array}{cc}
 -\frac{1}{4} & -\frac{3}{4} \\
 -\frac{3}{4} & -\frac{1}{4} \\
\end{array}
\right),
\end{equation}
\begin{equation}
c=(1/\sqrt{2},1/\sqrt{2}).
\end{equation}
With the same approach, we could obtain bases with higher $NL$ and different $J^P$ quantum numbers. In Table~\ref{tab:base}, we present the basis states for $ N \leq 3 $. Using these bases, the Schr\"odinger equation is transformed into an eigenvalue problem
\begin{equation}\label{eq:eigen}
\sum\limits_j H_{ij} \psi_j = M C_i \psi_i,
\end{equation}
where $ H_{ij} = \langle \psi_i | \hat{H} | \psi_j \rangle $ represents the matrix element of the Hamiltonian given in Eq.~(\ref{eq:H}). Here, $ M $ is the eigenvalue of Eq.~(\ref{eq:eigen}) and corresponds to the mass of the system. The coefficients $ C_i $ form the eigenvector, which can be used to construct the wave function as $ \Psi = \sum\limits_i C_i \psi_i $. Since the wave functions are fully anti-symmetric, the calculations could be simplified as
\begin{equation}
H_{ij} = \langle \psi_i | \hat{H} | \psi_j\rangle=3\langle \psi_i | T_3+V_{12} | \psi_j\rangle.
\end{equation}
Since in the frame of center-of-mass, ${\bf p}_\lambda=-{\bf p}_3$, we chose $T_3$ rather than $T_1$ or $T_2$. It is natural to use $V_{12}$ since the quarks 1 and 2 are exactly on the $\rho$-mode. In the potentials, except the simple central part, there are also complex spin-dependent interactions like hyperfine and spin-orbit terms. In Refs.~\cite{Capstick:1986ter,Murthy:1985fd}, the authors presented the analytical expressions. 

\begin{table}[htbp]
\caption{The parameters of the potential.}
\label{tab:parameter}
\renewcommand\arraystretch{1.5}
\begin{tabular*}{86mm}{@{\extracolsep{\fill}}lcccc}
\toprule[1.00pt]
\toprule[1.00pt]
Systems  &$\alpha_s$ &$b$~(${\rm GeV}^2$) &$\sigma$~(GeV) &$C$~(GeV) \\
\midrule[0.75pt]
$\Omega$ &0.660      &0.116               &1.600           &-0.539   \\
$\Xi$    &0.560      &0.122               &1.600           &-0.633   \\
Meson    &0.578      &0.144               &1.028           &-0.685   \\
\midrule[0.75pt]
\multicolumn{5}{c}{$m_{u/d}=0.370$ GeV~~~~~$m_{s}=0.600$ GeV}\\
\bottomrule[1.00pt]
\bottomrule[1.00pt]
\end{tabular*}
\end{table}

\begin{table*}[htbp]
\caption{Comparisons of $\Omega$ baryon masses (in MeV) across different theoretical and experimental studies.}
\label{tab:mass}
\renewcommand\arraystretch{1.5}
\begin{tabular*}{178mm}{@{\extracolsep{\fill}}cclcccccccc}
\toprule[1.00pt]
\toprule[1.00pt]
$|J^P\rangle_n$             &Dominate $N$ &Dominate Components                                                                &Our  &Exp.                                  &Ref.~\cite{Capstick:1986ter} &Ref.~\cite{Liu:2019wdr} &Ref.~\cite{Faustov:2015eba} &Ref.~\cite{Pervin:2007wa} &Ref.~\cite{Chao:1980em} &Ref.~\cite{Chen:2009de} \\
\midrule[0.75pt]
$|\frac{1}{2}^+\rangle_{1}$ &2            &$+0.941|\psi_{1/2^+}^{2,0,1/2}\rangle-0.338|\psi_{1/2^+}^{2,2,3/2}\rangle$         &2223 &$\cdots$                              &2220                         &2141                    &2301                        &2175                      &2190                    &2182                    \\
$|\frac{1}{2}^+\rangle_{2}$ &2            &$+0.345|\psi_{1/2^+}^{2,0,1/2}\rangle+0.938|\psi_{1/2^+}^{2,2,3/2}\rangle$         &2239 &$\cdots$                              &2255                         &2232                    &$\cdots$                    &2191                      &2210                    &2202                    \\
$|\frac{3}{2}^+\rangle_{1}$ &0            &$|\psi_{3/2^+}^{0,0,3/2}\rangle$                                                   &1672 &1672~\cite{ParticleDataGroup:2024cfk} &1635                         &1672                    &1678                        &1656                      &1675                    &1673                    \\
$|\frac{3}{2}^+\rangle_{2}$ &2            &$|\psi_{3/2^+}^{2,0,3/2}\rangle$                                                   &2164 &2109~\cite{BESIII:2024eqk}?           &2165                         &2159                    &2173                        &2170                      &2065                    &2078                    \\
$|\frac{3}{2}^+\rangle_{3}$ &2            &$+0.463|\psi_{3/2^+}^{2,2,1/2}\rangle-0.886|\psi_{3/2^+}^{2,2,3/2}\rangle$         &2232 &$\cdots$                              &2280                         &2188                    &2304                        &2182                      &2215                    &2208                    \\
$|\frac{3}{2}^+\rangle_{4}$ &2            &$+0.887|\psi_{3/2^+}^{2,2,1/2}\rangle+0.461|\psi_{3/2^+}^{2,2,3/2}\rangle$         &2284 &$\cdots$                              &2345                         &2245                    &2332                        &$\cdots$                  &2265                    &2263                    \\
$|\frac{5}{2}^+\rangle_{1}$ &2            &$-0.641|\psi_{5/2^+}^{2,2,1/2}\rangle+0.767|\psi_{5/2^+}^{2,2,3/2}\rangle$         &2216 &$\cdots$                              &2280                         &2252                    &2401                        &2178                      &2225                    &2224                    \\
$|\frac{5}{2}^+\rangle_{2}$ &2            &$-0.765|\psi_{5/2^+}^{2,2,1/2}\rangle-0.644|\psi_{5/2^+}^{2,2,3/2}\rangle$         &2283 &$\cdots$                              &2345                         &2303                    &$\cdots$                    &2210                      &2265                    &2260                    \\
$|\frac{7}{2}^+\rangle_{1}$ &2            &$|\psi_{7/2^+}^{2,2,3/2}\rangle$                                                   &2230 &$\cdots$                              &2295                         &2321                    &2369                        &2183                      &2210                    &2205                    \\
$|\frac{1}{2}^-\rangle_{1}$ &1            &$|\psi_{1/2^-}^{1,1,1/2}\rangle$                                                   &1957 &$\cdots$                              &1950                         &1957                    &1941                        &1923                      &2020                    &2015                    \\
$|\frac{1}{2}^-\rangle_{2}$ &3            &$|\psi_{1/2^-}^{3,1,1/2}\rangle_{2}$                                               &2384 &$\cdots$                              &2410                         &$\cdots$                &2463                        &$\cdots$                  &$\cdots$                &$\cdots$                \\
$|\frac{1}{2}^-\rangle_{3}$ &3            &$|\psi_{1/2^-}^{3,1,3/2}\rangle$                                                   &2446 &$\cdots$                              &2490                         &$\cdots$                &2580                        &$\cdots$                  &$\cdots$                &$\cdots$                \\
$|\frac{1}{2}^-\rangle_{4}$ &3            &$|\psi_{1/2^-}^{3,1,1/2}\rangle_{1}$                                               &2524 &$\cdots$                              &$\cdots$                     &$\cdots$                &$\cdots$                    &$\cdots$                  &$\cdots$                &$\cdots$                \\
$|\frac{3}{2}^-\rangle_{1}$ &1            &$|\psi_{3/2^-}^{1,1,1/2}\rangle$                                                   &2001 &2012~\cite{Belle:2018mqs}             &2000                         &2012                    &2038                        &1953                      &2020                    &2015                    \\
$|\frac{3}{2}^-\rangle_{2}$ &3            &$+0.932|\psi_{3/2^-}^{3,1,1/2}\rangle_{2}-0.362|\psi_{3/2^-}^{3,1,3/2}\rangle$     &2412 &$\cdots$                              &2440                         &$\cdots$                &2537                        &$\cdots$                  &$\cdots$                &$\cdots$                \\
$|\frac{3}{2}^-\rangle_{3}$ &3            &$-0.305|\psi_{3/2^-}^{3,1,1/2}\rangle_{1}+0.340|\psi_{3/2^-}^{3,1,1/2}\rangle_{2}$ &2437 &$\cdots$                              &2495                         &$\cdots$                &2636                        &$\cdots$                  &$\cdots$                &$\cdots$                \\
                            &             &$+0.890|\psi_{3/2^-}^{3,1,3/2}\rangle$                                             &     &                                      &                             &                        &                            &                          &                        &                        \\
\addlinespace[0.55em]
$|\frac{3}{2}^-\rangle_{4}$ &3            &$-0.692|\psi_{3/2^-}^{3,1,1/2}\rangle_{1}-0.386|\psi_{3/2^-}^{3,2,1/2}\rangle$     &2494 &$\cdots$                              &$\cdots$                     &$\cdots$                &$\cdots$                    &$\cdots$                  &$\cdots$                &$\cdots$                \\
                            &             &$+0.611|\psi_{3/2^-}^{3,3,3/2}\rangle$                                             &     &                                      &                             &                        &                            &                          &                        &                        \\
\addlinespace[0.55em]
$|\frac{3}{2}^-\rangle_{5}$ &3            &$+0.662|\psi_{3/2^-}^{3,1,1/2}\rangle_{1}+0.750|\psi_{3/2^-}^{3,3,3/2}\rangle$     &2514 &$\cdots$                              &$\cdots$                     &$\cdots$                &$\cdots$                    &$\cdots$                  &$\cdots$                &$\cdots$                \\
$|\frac{3}{2}^-\rangle_{6}$ &3            &$-0.942|\psi_{3/2^-}^{3,2,1/2}\rangle-0.337|\psi_{3/2^-}^{3,3,3/2}\rangle$         &2570 &$\cdots$                              &$\cdots$                     &$\cdots$                &$\cdots$                    &$\cdots$                  &$\cdots$                &$\cdots$                \\
$|\frac{5}{2}^-\rangle_{1}$ &3            &$|\psi_{5/2^-}^{3,1,3/2}\rangle$                                                   &2431 &$\cdots$                              &2490                         &$\cdots$                &2653                        &$\cdots$                  &$\cdots$                &$\cdots$                \\
$|\frac{5}{2}^-\rangle_{2}$ &3            &$|\psi_{5/2^-}^{3,3,1/2}\rangle$                                                   &2445 &$\cdots$                              &$\cdots$                     &$\cdots$                &$\cdots$                    &$\cdots$                  &$\cdots$                &$\cdots$                \\
$|\frac{5}{2}^-\rangle_{3}$ &3            &$-0.590|\psi_{5/2^-}^{3,2,1/2}\rangle+0.807|\psi_{5/2^-}^{3,3,3/2}\rangle$         &2503 &$\cdots$                              &$\cdots$                     &$\cdots$                &$\cdots$                    &$\cdots$                  &$\cdots$                &$\cdots$                \\
$|\frac{5}{2}^-\rangle_{4}$ &3            &$+0.797|\psi_{5/2^-}^{3,2,1/2}\rangle+0.604|\psi_{5/2^-}^{3,3,3/2}\rangle$         &2536 &$\cdots$                              &$\cdots$                     &$\cdots$                &$\cdots$                    &$\cdots$                  &$\cdots$                &$\cdots$                \\
$|\frac{7}{2}^-\rangle_{1}$ &3            &$-0.856|\psi_{7/2^-}^{3,3,1/2}\rangle-0.517|\psi_{7/2^-}^{3,3,3/2}\rangle$         &2478 &$\cdots$                              &$\cdots$                     &$\cdots$                &2599                        &$\cdots$                  &$\cdots$                &$\cdots$                \\
$|\frac{7}{2}^-\rangle_{2}$ &3            &$-0.522|\psi_{7/2^-}^{3,3,1/2}\rangle+0.853|\psi_{7/2^-}^{3,3,3/2}\rangle$         &2515 &$\cdots$                              &$\cdots$                     &$\cdots$                &$\cdots$                    &$\cdots$                  &$\cdots$                &$\cdots$                \\
$|\frac{9}{2}^-\rangle_{1}$ &3            &$|\psi_{9/2^-}^{3,3,3/2}\rangle$                                                   &2485 &$\cdots$                              &$\cdots$                     &$\cdots$                &2649                        &$\cdots$                  &$\cdots$                &$\cdots$                \\
\bottomrule[1.00pt]
\bottomrule[1.00pt]
\end{tabular*}
\end{table*}

\begin{table}
\centering
\caption{The $\beta$ values (in GeV) used in this work. $\tilde{\Xi}$ implies that the state is $\rho$-mode excited.}
\label{tab:beta}
\renewcommand\arraystretch{1.35}
\begin{tabular*}{86mm}{@{\extracolsep{\fill}}lcclcclc}
\toprule[1.00pt]
\toprule[1.00pt]
States                          &$\beta_\rho$ &$\beta_\lambda$ &States                            &$\beta_\rho$ &$\beta_\lambda$ &States           &$\beta$ \\
\midrule[0.75pt]
$\Xi(1S,\frac{1}{2}^+)$         &0.270        &0.297           &$\Xi^*(1S,\frac{1}{2}^+)$         &0.246       &0.253           &$\pi$            &0.409   \\
$\tilde{\Xi}(2S,\frac{1}{2}^+)$ &0.161        &0.211           &$\tilde{\Xi}^*(2S,\frac{1}{2}^+)$ &0.142       &0.218           &$K$              &0.385   \\
$\Xi(2S,\frac{1}{2}^+)$         &0.197        &0.159           &$\Xi^*(2S,\frac{3}{2}^+)$         &0.220       &0.144           &$\eta(s\bar{s})$ &0.403   \\
$\Xi(1P)$                       &0.235        &0.186           &$\tilde{\Xi}(1P)$                 &0.178       &0.249           &                 &        \\
$\Xi(1D)$                       &0.230        &0.143           &$\tilde{\Xi}(1D)$                 &0.141       &0.234           &                 &        \\
$\hat{\Xi}(1D)$                 &0.164        &0.182           &$\Omega$                          &0.260       &0.302           &                 &        \\
\midrule[0.75pt]
\bottomrule[1.00pt]
\bottomrule[1.00pt]
\end{tabular*}
\end{table}

Although a state with the same $ J^P $ may have many components $ |\psi_{J^P}^{NLS}\rangle $, we observe that by selecting specific values of $\beta$, one or a few basis states dominate the wave function. This allows us to describe the state using as few basis components as possible. In this work, we adopt $\beta_\rho = 0.260$. Combining the observed $\Omega$, $\Xi$, and light flavor mesons, we obtain the parameters of the potential, which are listed in Table~\ref{tab:parameter}. In addition, the spectra of the $\Omega$ are presented in Table~\ref{tab:mass}. 

For $\Omega$ states with the same $ J^P $, increasing $ N $ by 2 leads to mass increases of several hundred MeV. These large mass gaps indicate that mixing between states with different $ N $ is minimal. Consequently, the solutions are dominated by a single $ N $, and contributions from other $ N $ values can be neglected. However, for states with the same $ N $, mixing effects may not be negligible. To address this, we select the largest components and normalize the wave functions accordingly.

According to Table~\ref{tab:mass}, we could obtain:
\begin{enumerate}
\item The mass of the ground state $\Omega$ is 1672 MeV, which serves as a reliable anchor point for determining the parameters of the potential model.
\item The $\Omega(2012)$ is a strong candidate for the $1P$ state with $J^P=3/2^-$. And the mass of $\Omega(1P,1/2^-)$ is calculate as about 1.95 GeV.
\item The mass of the newly observed $\Omega(2109)$ is much higher than the $1P$ states and much less than the $N=3$ states. For the $N=2$ states, the lowest state is $\Omega(2S,3/2^+)$, which is the only state may match the $\Omega(2109)$ in our scheme. However, the mass of $\Omega(2109)$ is still about 50 MeV below the calculation of $\Omega(2S,3/2^+)$. But in some works~\cite{Chao:1980em,Chen:2009de}, lower masses of $\Omega(2S,3/2^+)$ are obtained. One possible reason is that the potential especially the confinements may be different from our calculations. In Sec.~\ref{sec:Decay_Behavior}, we investigate the decay behaviors of $\Omega(2S,3/2^+)$, which provides more clues in analysis of $\Omega(2109)$.
\item For most $ N=2 $ states, the calculated masses are around 2.2 GeV, making the observed $\Omega(2280)$ a good candidate for this category.
\item Theoretical masses for most $ N=3 $ states fall within the range of 2.38-2.57 GeV, which includes $\Omega(2380)$ and $\Omega(2470)$. Thus, $\Omega(2380)$ and $\Omega(2470)$ are likely candidate for the $ N=3 $ states.
\end{enumerate}

\begin{table}[htbp]
\caption{Calculated strong decay widths (in MeV) for the $N=1$ states. The number ``0.0" imply that the width is less then 0.1 MeV.}
\label{tab:widthsN1}
\renewcommand\arraystretch{1.5}
\begin{tabular*}{86mm}{@{\extracolsep{\fill}}lcccc}
\toprule[1.00pt]
\toprule[1.00pt]
                    &$M_b^f$ (MeV) &$\Omega|\frac{1}{2}^-\rangle_{1}$ &$\Omega|\frac{3}{2}^-\rangle_{1}$ \\
$M_b^i$ (MeV)       &              &1957                              &2001                              \\
\midrule[0.75pt]
$\Xi(1S)\bar{K}$    &1314          &21.9                              &5.8                               \\
$\Xi(1S)\pi\bar{K}$ &1314          &0.0                               &0.3                               \\
\midrule[0.75pt]
Total               &              &21.9                              &6.1                                \\
\bottomrule[1.00pt]
\bottomrule[1.00pt]
\end{tabular*}
\end{table}

\begin{table*}[htbp]
\caption{Calculated strong decay widths (in MeV) for the $N=2$ states. The $M_b^f$ and $M_b^i$ are masses of final and initial baryons. The symbol ``$\cdots$" imply that the state locates below the threshold of the channel. The number ``0.0" imply that the width is less then 0.1 MeV.}
\label{tab:widthsN2}
\renewcommand\arraystretch{1.5}
\centering
\begin{tabular*}{180mm}{@{\extracolsep{\fill}}lcccccccccc}
\toprule[1.00pt]
\toprule[1.00pt]
                                      &$M_b^f$ (MeV) &$\Omega|\frac{1}{2}^+\rangle_{1}$ &$\Omega|\frac{1}{2}^+\rangle_{2}$ &$\Omega|\frac{3}{2}^+\rangle_{2}$ &$\Omega|\frac{3}{2}^+\rangle_{2}$ &$\Omega|\frac{3}{2}^+\rangle_{3}$ &$\Omega|\frac{3}{2}^+\rangle_{4}$ &$\Omega|\frac{5}{2}^+\rangle_{1}$ &$\Omega|\frac{5}{2}^+\rangle_{2}$ &$\Omega|\frac{7}{2}^+\rangle_{1}$ \\
$M_b^i$ (MeV)                         &              &2223                              &2239                              &2164                              &2109                              &2232                              &2284                              &2216                              &2283                              &2230                              \\
\midrule[0.75pt]
$\Xi(1S)\bar{K}$                      &1314          &2.7                               &53.0                              &8.1                               &15.0                              &15.0                              &12.5                              &13.4                              &0.3                               &46.8                              \\
$\Xi^*(1S)\bar{K}$                    &1535          &34.1                              &1.1                               &46.7                              &29.3                              &20.2                              &19.7                              &8.1                               &54.2                              &5.0                               \\
$\Xi_{1}(1P,\frac{3}{2}^-)\bar{K}$    &1775          &$\cdots$                          &$\cdots$                          &$\cdots$                          &$\cdots$                          &$\cdots$                          &27.7                              &$\cdots$                          &0.0                               &$\cdots$                          \\
$\Xi(1S)\bar{K}^*$                    &1314          &0.0                               &26.4                              &$\cdots$                          &$\cdots$                          &4.4                               &19.7                              &0.0                               &0.0                               &0.0                               \\
$\Omega|\frac{3}{2}^+\rangle_{1}\eta$ &1672          &0.2                               &0.0                               &$\cdots$                          &$\cdots$                          &0.5                               &6.2                               &$\cdots$                          &19.8                              &0.0                               \\
Small widths                          &              &0.0                               &0.0                               &0.0                               &0.0                               &0.0                               &0.1                               &0.0                               &0.5                               &0.0                               \\
\midrule[0.75pt]
Total                                 &              &37.0                              &80.5                              &54.8                              &44.3                              &40.1                              &85.9                              &21.5                              &74.8                              &51.8                              \\
\bottomrule[1.00pt]
\bottomrule[1.00pt]
\end{tabular*}
\begin{flushleft}
Without color and flavor part, the spin-spatial wave functions of $\Xi(1P)$ are defined as
\begin{equation}\nonumber
\left(\begin{array}{c}
|\Xi_1(1P,\frac{1}{2}^-)\rangle\\
|\Xi_2(1P,\frac{1}{2}^-)\rangle\\
|\Xi_3(1P,\frac{1}{2}^-)\rangle
\end{array}\right)=\left(
\begin{array}{rrr}
  0.306291 & -0.085464 & -0.948094 \\
  0.903305 &  0.340361 &  0.261140 \\
  0.300376 & -0.936403 &  0.181450 \\
\end{array}
\right)
\left(\begin{array}{c}
|[[R_{00}^\rho R_{01}^\lambda]_{1}\chi_{1,\frac{1}{2}}]_{\frac{1}{2}}\rangle\\
|[[R_{00}^\rho R_{01}^\lambda]_{1}\chi_{1,\frac{3}{2}}]_{\frac{1}{2}}\rangle\\
|[[R_{01}^\rho R_{00}^\lambda]_{1}\chi_{0,\frac{1}{2}}]_{\frac{1}{2}}\rangle\\
\end{array}\right),
\end{equation}
\begin{equation}\nonumber
\left(\begin{array}{c}
|\Xi_1(1P,\frac{3}{2}^-)\rangle\\
|\Xi_2(1P,\frac{3}{2}^-)\rangle\\
|\Xi_3(1P,\frac{3}{2}^-)\rangle
\end{array}\right)=\left(
\begin{array}{rrr}
 -0.491418 & -0.373175 &  0.786923 \\
  0.554941 &  0.562207 &  0.613159 \\
  0.671229 & -0.738013 &  0.069189 \\
\end{array}
\right)
\left(\begin{array}{c}
|[[R_{00}^\rho R_{01}^\lambda]_{1}\chi_{1,\frac{1}{2}}]_{\frac{3}{2}}\rangle\\
|[[R_{00}^\rho R_{01}^\lambda]_{1}\chi_{1,\frac{3}{2}}]_{\frac{3}{2}}\rangle\\
|[[R_{01}^\rho R_{00}^\lambda]_{1}\chi_{0,\frac{1}{2}}]_{\frac{3}{2}}\rangle\\
\end{array}\right).
\end{equation}
\end{flushleft}
\end{table*}

\section{Strong decay properties}\label{sec:Decay_Behavior}
In addition to mass, the decay width is a crucial property of hadrons. In this work, we employ the Quark Pair Creation (QPC) model~\cite{Micu:1968mk,LeYaouanc:1973ldf,Zhao:2017fov,Chen:2017gnu,Chen:2016iyi,Chen:2007xf,Lu:2018utx,Wang:2018hmi} to calculate the decay widths of the $\Omega$ baryons. The transition operator in the QPC model is given by:
\begin{equation}\label{eq:toperator}
\begin{split}
\hat{\cal T} = &-3\gamma \sum_{m} C_{1,m;1,-m}^{0,0} \int {\rm d}^3{\bf p}_i \, {\rm d}^3{\bf p}_j \, \delta({\bf p}_i + {\bf p}_j) \\
&\times \mathcal{Y}_1^m\left(\frac{{\bf p}_i - {\bf p}_j}{2}\right) \omega_0 \phi_0 \chi_{1,-m} b^\dagger_i({\bf p}_i) d^\dagger_j({\bf p}_j),
\end{split}
\end{equation}
where $\omega_0$, $\phi_0$, $\chi_{1,-m}$, and $\mathcal{Y}_1^m$ represent the color, flavor, spin, and spatial wave functions of the created quark pair, respectively. The parameter $\gamma$ describes the strength of quark-pair creation from the vacuum. 

For a decay process $A \to BC$ with relative angular momentum $L$ and total spin $S$ of the final state $BC$, the partial wave amplitude is:
\begin{equation}\label{eq:MAtoBC}
{\cal M}_{A\to BC}^{LS}(p) = \langle \Psi_B \Psi_C, LS, p | \hat{\cal T} | \Psi_A \rangle,
\end{equation}
where $p$ is the momentum of hadron $B$. The decay width of $A \to BC$ is then calculated as
\begin{equation}
\Gamma_{A\to BC} = 2\pi \frac{E_B(p) E_C(p)}{M_A} p \sum\limits_{LS} |{\cal M}_{A\to BC}^{LS}(p)|^2,
\end{equation}
where $E_B(p) = \sqrt{M_B^2 + p^2}$ and $E_C(p) = \sqrt{M_C^2 + p^2}$ are the energies of the final hadrons $B$ and $C$, respectively.

Considering that the final state may contain a $\Xi$ baryon and a light flavor meson, we also utilize the simple harmonic oscillator wave functions for the spatial parts of $\Xi$ and light flavor mesons when calculating the amplitude of the QPC model. With the approach in Refs.~\cite{Chen:2018orb,Luo:2023sra,Luo:2023sne}, we extract the $\beta$ values of the harmonic oscillator wave functions from the potential model. In Table~\ref{tab:beta}, we present these $\beta$ values, which are employed to the latter calculations. The remaining parameter of is the $\gamma$ value. We notice that the $\Xi^*\to \Xi\pi$ has similar decay mode of the excited $\Omega$, we determine $\gamma=8.28$ via the width of $\Xi^*$.

\subsection{$N=1$ states}

The $ N = 1 $ states correspond to the $\Omega(1P)$ multiplet, which includes $\Omega(1P, 1/2^-)$ and $\Omega(1P, 3/2^-)$. Their calculated decay widths are presented in Table~\ref{tab:widthsN1}. As previously discussed, $\Omega(2012)$ is a strong candidate for the $ N = 1 $ state. The Belle Collaboration~\cite{Belle:2018mqs} reports the mass and width of $\Omega(2012)$ as:
\begin{equation}
\begin{split}
m &= 2012.4 \pm 0.7 \pm 0.6 \, \text{MeV}, \\
\Gamma &= 6.4_{-2.0}^{+2.5} \pm 1.6 \, \text{MeV}.
\end{split}
\end{equation}
In our calculations, the predicted width of the $ J^P = 3/2^- $ state is 6.1 MeV, which aligns well with the measured value for $\Omega(2012)$. Similar theoretical results have been obtained in Refs.~\cite{Xiao:2018pwe,Wang:2018hmi,Liu:2019wdr,Aliev:2018yjo,Zhong:2022cjx}. Thus the $\Omega$ is a good candidate of $\Omega(1P, 3/2^-)$.

The other $ N = 1 $ state, $\Omega(1P, 1/2^-)$, has a predicted width of approximately 21.9 MeV. Based on spin-parity considerations, $\Omega(1P, 1/2^-)$ decays into $\Xi \bar{K}$ via an $ S $-wave, while $\Omega(1P, 3/2^-)$ couples to $\Xi \bar{K}$ in a $ D $-wave. In addition to $\Omega(2012)$, there may be a missing $\Omega$ state around 1.95 GeV. We encourage experimental efforts to search for this state.

Besides the two-body decay process, we also investigate the $\Omega(2012)$ via the three-body decay. For the $\Omega(2012)$, the $\Xi^*\bar{K}$ is kinematically forbidden when we take the central masses of $\Omega(2012)$ and $\Xi^*$. But $\Xi\pi\bar{K}$ is allowed. In 2022, the Belle Collaboration measured the branching ratio of $\Xi \pi \bar{K}$ to $\Xi \bar{K}$ and obtained~\cite{Belle:2022mrg}\footnote{
In 2019, the Belle Collaboration reported the result with ${\cal R}^{\Xi \pi \bar{K}}_{\Xi \bar{K}}<11.9\%$~\cite{Belle:2019zco}. However, this analysis was corrected by the Belle Collaboration in 2022 as indicated in Ref.~\cite{Belle:2022mrg}.
}
\begin{equation}
{\cal R}^{\Xi \pi \bar{K}}_{\Xi \bar{K}}=\frac{{\cal B}(\Omega(2012)\to \Xi(1530)(\to \Xi\pi)\bar{K})}{{\cal B}(\Omega(2012)\to \Xi \bar{K})}=0.99\pm0.26\pm0.06.
\end{equation}
Based on this result, we also perform a width of the three-body decay of $\Omega(1P)$. For a three-body decay $A\to RD\to BCD$, the decay width could be obtained with the following approach~\cite{Luo:2009wu,Yu:2011ta}:
\begin{equation}
\Gamma_{A\to RD\to BCD}\approx\int_{(m_B+m_C)^2}^{(m_A-m_D)^2}{\rm d}s_R\frac{\sqrt{s_R}}{\pi}\frac{\Gamma_{A\to RD}(s_R)\Gamma_{R\to AB}(s_R)}{(s_R-m_R^2)^2+(m_R\Gamma_R)^2},
\end{equation}
where the $m_R$ and $\Gamma_R$ are Breit-Wigner mass and width of the intermediate state $R$. For the $\Omega(1P)$, we select the $\Xi^*$ as the intermediate, then the three-body decays are $\Omega(1P)\to \Xi^*\bar{K}\to \Xi\pi\bar{K}$. The numerical results are presented in Table~\ref{tab:widthsN1}. As shown in Table~\ref{tab:widthsN1}, the partial width of $\Omega(1P,1/2^-)\to \Xi\pi\bar{K}$ is very small. On the one hand, the phase space of this process is very small. On the other hand, the coupling between $\Omega(1P,1/2^-)$ and $ \Xi^*\bar{K}$ is $D$-wave, the strength of which is highly depressed. For the $\Omega(1P,3/2^-)$, the partial width of $\Xi\pi\bar{K}$ channel is calculated as 0.3 MeV. In our scheme, the relative branching ratio is calculated as ${\cal R}_{\Xi\bar{K}}^{\Xi\pi\bar{K}}\approx0.05$. This result is similar to the quark model calculation in Refs.~\cite{Zhong:2022cjx,Arifi:2022ntc} but far from the measurement by the Belle Collaboration in 2022~\cite{Belle:2022mrg}. This presents us more options for assignments of the $\Omega(2012)$. Since the $\Omega(2012)$ is close to the $\Xi^*\bar{K}$ threshold, it is also could be interpreted as hadronic molecular candidates~\cite{Valderrama:2018bmv,Pavao:2018xub,Ikeno:2020vqv,Gutsche:2019eoh}, where the $\Xi\pi\bar{K}$ channel may have larger branching ratio than that of the $\Omega(1P,3/2^-)$ state.

\subsection{$N=2$ states}

The $N=2$ $\Omega$ excited states contain $2S$, $1D$, and $2S$-$1D$ mixing schemes, which could be referred from the basis in Table~\ref{tab:base}. The calculated widths are presented in Table~\ref{tab:widthsN2}. For the $\Omega(2S,3/2^+)$, which is labeled as $\Omega|\frac{3}{2}^+\rangle_2$, if we take the mass with 2164 MeV, the obtained width is 54.8 MeV. We find that this value is much larger than the central width of the measured result of $\Omega(2109)$. Then if we take the mass with 2109 MeV, the calculated width is 44.3 MeV. This value is also higher than the central value of the measured width of $\Omega(2109)$. However, if we consider the measurement error, the theoretical width could match the experimental result. For $\Omega(2109)$, the most obvious controversy is the mass. On the one hand, as discussed in Sec.~\ref{sec:Mass_Spectrum}, if we change the form of confinement potential, the mass of $\Omega(2S,3/2^+)$ may could be depressed. On the other hand, we suggest search for excited $\Omega$ state around 2.16 GeV to verify the calculations.

The masses of the remaining $N=2$ $\Omega$ states are in the region of 2.2$\sim$2.3 GeV. In PDG~\cite{ParticleDataGroup:2024cfk}, the state matched this region is $\Omega(2250)$. However, decoding $\Omega(2250)$ among these states in Table~\ref{tab:widthsN2} is very difficult. Considering that $\Omega(2250)$ was observed in $\Xi^*\bar{K}$ channel~\cite{Biagi:1985rn,Aston:1987bb}, the $\Xi^*\bar{K}$ channel should occupy a significant partial width. According to Table~\ref{tab:widthsN2}, the calculated width of $\Omega|1/2^+\rangle_2\to \Xi^*\bar{K}$ is only about 1.1 MeV. Thus it is difficult to assign the $\Omega(2250)$ to $\Omega|1/2^+\rangle_2$. According to Table~\ref{tab:widthsN2}, the $\Omega|\frac{1}{2}^+\rangle_1$, $\Omega|\frac{3}{2}^+\rangle_3$, $\Omega|\frac{5}{2}^+\rangle_2$ etc. occupy large $\Xi^*\bar{K}$ branching ratio. On the other hand, most of these state also have considerable $\Xi\bar{K}$ partial width. Thus we suggest to measure  branching ratios of $\Gamma_{\Xi\bar{K}}:\Gamma_{\Xi^*\bar{K}}:\Gamma_{\Xi\bar{K}^*}$ of $\Omega(2250)$, which could provide crucial clues to decode the $\Omega(2250)$.

For the most $N=2$ $\Omega$ states, $\Xi\bar{K}$ and $\Xi^*\bar{K}$ are two important channels. For some states, the partial width of $\Xi\bar{K}^*$ channel also occupy large branching ration. In 2.2$\sim$2.3 GeV, besides $\Omega(2109)$ and $\Omega(2250)$, there exist a series of excited $\Omega$ states. Except $\Omega|1/2^+\rangle_2$ and $\Omega|3/2^+\rangle_4$, the widths of most states are calculated to be less than 60 MeV. It is potential to observe these states.

\subsection{$N=3$ states}

\begin{table*}[htbp]
\caption{Calculated strong decay widths (in MeV) for the $N=3$ states. The conventions are the same as Table~\ref{tab:widthsN2}.}
\label{tab:widthsN3}
\renewcommand\arraystretch{1.35}
\begin{tabular*}{180mm}{@{\extracolsep{\fill}}lccccccccc}
\toprule[1.00pt]
\toprule[1.00pt]
                                        &$M_b^f$ (MeV) &$\Omega|\frac{1}{2}^-\rangle_{2}$ &$\Omega|\frac{1}{2}^-\rangle_{3}$ &$\Omega|\frac{1}{2}^-\rangle_{4}$ &$\Omega|\frac{3}{2}^-\rangle_{2}$ &$\Omega|\frac{3}{2}^-\rangle_{3}$ &$\Omega|\frac{3}{2}^-\rangle_{4}$ &$\Omega|\frac{3}{2}^-\rangle_{5}$ &$\Omega|\frac{3}{2}^-\rangle_{6}$ \\
$M_b^i$ (MeV)                           &              &2384                              &2446                              &2524                              &2412                              &2437                              &2494                              &2514                              &2570                              \\
\midrule[0.75pt]
$\Xi(1S)\bar{K}$                        &1314          &3.0                               &13.4                              &0.0                               &0.0                               &0.1                               &4.2                               &4.1                               &0.6                               \\
$\tilde{\Xi}(2S)\bar{K}$                &1902          &$\cdots$                          &34.6                              &0.8                               &0.0                               &0.1                               &0.5                               &0.1                               &0.2                               \\
$\Xi^*(1S)\bar{K}$                      &1535          &26.6                              &0.7                               &0.6                               &19.3                              &17.2                              &2.4                               &1.1                               &0.3                               \\
$\Xi_{1}(1P,\frac{1}{2}^-)\bar{K}$      &1746          &0.6                               &1.1                               &36.6                              &35.0                              &5.9                               &6.2                               &2.3                               &17.4                              \\
$\Xi_{2}(1P,\frac{1}{2}^-)\bar{K}$      &1784          &0.6                               &7.4                               &1.2                               &2.7                               &6.0                               &7.6                               &1.0                               &0.5                               \\
$\Xi_{3}(1P,\frac{1}{2}^-)\bar{K}$      &1811          &0.0                               &8.1                               &3.3                               &0.2                               &0.0                               &2.1                               &3.2                               &0.1                               \\
$\Xi_{1}(1P,\frac{3}{2}^-)\bar{K}$      &1775          &37.1                              &0.0                               &3.4                               &21.4                              &13.3                              &3.1                               &11.6                              &13.3                              \\
$\Xi_{2}(1P,\frac{3}{2}^-)\bar{K}$      &1825          &12.8                              &0.2                               &3.2                               &3.5                               &1.5                               &2.7                               &3.4                               &7.6                               \\
$\Xi_{3}(1P,\frac{3}{2}^-)\bar{K}$      &1880          &1.0                               &11.2                              &0.1                               &0.3                               &6.3                               &2.1                               &2.7                               &1.3                               \\
$\Xi(1P,\frac{5}{2}^-)\bar{K}$          &1840          &0.4                               &0.4                               &0.1                               &21.8                              &26.2                              &0.8                               &0.3                               &0.2                               \\
$\Xi_{1}(1D,\frac{3}{2}^+)\bar{K}$      &2015          &$\cdots$                          &$\cdots$                          &0.0                               &$\cdots$                          &$\cdots$                          &$\cdots$                          &6.2                               &1.1                               \\
$\Xi_{2}(1D,\frac{3}{2}^+)\bar{K}$      &2058          &$\cdots$                          &$\cdots$                          &$\cdots$                          &$\cdots$                          &$\cdots$                          &$\cdots$                          &$\cdots$                          &10.1                              \\
$\Xi(1S)\bar{K}^*$                      &1314          &0.3                               &11.2                              &0.0                               &10.3                              &1.3                               &8.9                               &10.9                              &2.0                               \\
$\Xi^*(1S)\bar{K}^*$                    &1535          &$\cdots$                          &0.1                               &0.1                               &$\cdots$                          &33.4                              &0.6                               &0.7                               &0.4                               \\
$\Omega|\frac{3}{2}^+\rangle_{1}\eta$   &1672          &7.5                               &0.9                               &0.0                               &6.9                               &3.2                               &1.3                               &2.0                               &0.5                               \\
$\Omega|\frac{1}{2}^-\rangle_{1}\eta$   &1957          &$\cdots$                          &$\cdots$                          &2.5                               &$\cdots$                          &$\cdots$                          &$\cdots$                          &0.2                               &2.9                               \\
$\Omega|\frac{3}{2}^-\rangle_{1}\eta$   &2001          &$\cdots$                          &$\cdots$                          &$\cdots$                          &$\cdots$                          &$\cdots$                          &$\cdots$                          &$\cdots$                          &1.8                               \\
$\Omega|\frac{3}{2}^+\rangle_{1}\sigma$ &1672          &1.4                               &0.7                               &0.0                               &3.3                               &0.3                               &1.4                               &2.7                               &0.9                               \\
Small widths                            &              &0.0                               &0.0                               &0.1                               &0.0                               &0.0                               &0.1                               &0.4                               &0.3                               \\
\midrule[0.75pt]
Total                                   &              &91.3                              &90.0                              &52.0                              &124.7                             &114.7                             &44.0                              &52.9                              &61.5                              \\
\midrule[0.75pt]
\midrule[0.75pt]
                                        &$M_b^f$ (MeV) &$\Omega|\frac{5}{2}^-\rangle_{1}$ &$\Omega|\frac{5}{2}^-\rangle_{2}$ &$\Omega|\frac{5}{2}^-\rangle_{3}$ &$\Omega|\frac{5}{2}^-\rangle_{4}$ &$\Omega|\frac{7}{2}^-\rangle_{1}$ &$\Omega|\frac{7}{2}^-\rangle_{2}$ &$\Omega|\frac{9}{2}^-\rangle_{1}$ \\
$M_b^i$ (MeV)                           &              &2431                              &2445                              &2503                              &2536                              &2478                              &2515                              &2485                              \\
\midrule[0.75pt]
$\Xi(1S)\bar{K}$                        &1314          &0.1                               &2.7                               &2.6                               &1.1                               &14.6                              &0.8                               &29.7                              \\
$\Xi^*(1S)\bar{K}$                      &1535          &3.5                               &20.8                              &8.6                               &5.0                               &20.9                              &20.5                              &8.4                               \\
$\Xi_{1}(1P,\frac{1}{2}^-)\bar{K}$      &1746          &0.5                               &0.1                               &3.1                               &2.6                               &2.7                               &0.3                               &0.0                               \\
$\Xi_{2}(1P,\frac{1}{2}^-)\bar{K}$      &1784          &1.4                               &0.0                               &0.1                               &0.9                               &0.2                               &3.1                               &0.1                               \\
$\Xi_{3}(1P,\frac{1}{2}^-)\bar{K}$      &1811          &0.5                               &0.1                               &1.0                               &1.8                               &0.1                               &0.0                               &0.0                               \\
$\Xi_{1}(1P,\frac{3}{2}^-)\bar{K}$      &1775          &0.1                               &9.5                               &6.0                               &5.2                               &1.4                               &2.1                               &0.2                               \\
$\Xi_{2}(1P,\frac{3}{2}^-)\bar{K}$      &1825          &0.1                               &3.7                               &0.5                               &4.8                               &0.0                               &0.1                               &0.2                               \\
$\Xi_{3}(1P,\frac{3}{2}^-)\bar{K}$      &1880          &0.7                               &3.7                               &8.0                               &5.1                               &0.1                               &0.6                               &0.2                               \\
$\Xi(1P,\frac{5}{2}^-)\bar{K}$          &1840          &5.2                               &1.6                               &1.8                               &1.0                               &5.9                               &20.5                              &1.4                               \\
$\Xi_{1}(1D,\frac{5}{2}^+)\bar{K}$      &1974          &$\cdots$                          &$\cdots$                          &11.3                              &0.4                               &0.0                               &0.0                               &0.0                               \\
$\Xi(1S)\bar{K}^*$                      &1314          &29.6                              &4.3                               &6.2                               &3.3                               &3.8                               &0.3                               &8.3                               \\
$\Xi^*(1S)\bar{K}^*$                    &1535          &0.0                               &0.0                               &1.6                               &1.8                               &0.5                               &8.7                               &0.0                               \\
$\Omega|\frac{3}{2}^+\rangle_{1}\eta$   &1672          &3.6                               &3.4                               &5.0                               &3.2                               &4.5                               &19.9                              &0.7                               \\
$\Omega|\frac{3}{2}^+\rangle_{1}\sigma$ &1672          &3.8                               &1.4                               &1.0                               &0.7                               &2.3                               &1.5                               &1.7                               \\
Small widths                            &              &0.3                               &0.0                               &0.2                               &0.2                               &0.0                               &0.0                               &0.0                               \\
\midrule[0.75pt]
Total                                   &              &49.4                              &51.3                              &57.0                              &37.1                              &57.0                              &78.4                              &50.9                              \\
\bottomrule[1.00pt]
\bottomrule[1.00pt]
\end{tabular*}
\begin{flushleft}
The spin-spatial parts of the $\Xi(1D)$ used in this Table are
\begin{equation}\nonumber
\begin{split}
|\Xi_1(1D,3/2^+)\rangle=&
-0.206334|[[R_{00}^\rho R_{02}^\lambda]_{2}\chi_{1,\frac{1}{2}}]_{\frac{3}{2}}\rangle
-0.112008|[[R_{00}^\rho R_{02}^\lambda]_{2}\chi_{1,\frac{3}{2}}]_{\frac{3}{2}}\rangle
+0.037795|[[R_{01}^\rho R_{01}^\lambda]_{1}\chi_{0,\frac{1}{2}}]_{\frac{3}{2}}\rangle\\&
+0.063245|[[R_{01}^\rho R_{01}^\lambda]_{2}\chi_{0,\frac{1}{2}}]_{\frac{3}{2}}\rangle
+0.792941|[[R_{02}^\rho R_{00}^\lambda]_{2}\chi_{1,\frac{1}{2}}]_{\frac{3}{2}}\rangle
+0.557401|[[R_{02}^\rho R_{00}^\lambda]_{2}\chi_{1,\frac{3}{2}}]_{\frac{3}{2}}\rangle,
\end{split}
\end{equation}
\begin{equation}\nonumber
\begin{split}
|\Xi_2(1D,3/2^+)\rangle=&
-0.673552|[[R_{00}^\rho R_{02}^\lambda]_{2}\chi_{1,\frac{1}{2}}]_{\frac{3}{2}}\rangle
-0.306976|[[R_{00}^\rho R_{02}^\lambda]_{2}\chi_{1,\frac{3}{2}}]_{\frac{3}{2}}\rangle
-0.031699|[[R_{01}^\rho R_{01}^\lambda]_{1}\chi_{0,\frac{1}{2}}]_{\frac{3}{2}}\rangle\\&
+0.012292|[[R_{01}^\rho R_{01}^\lambda]_{2}\chi_{0,\frac{1}{2}}]_{\frac{3}{2}}\rangle
+0.226328|[[R_{02}^\rho R_{00}^\lambda]_{2}\chi_{1,\frac{1}{2}}]_{\frac{3}{2}}\rangle
-0.632229|[[R_{02}^\rho R_{00}^\lambda]_{2}\chi_{1,\frac{3}{2}}]_{\frac{3}{2}}\rangle,
\end{split}
\end{equation}
\begin{equation}\nonumber
\begin{split}
|\Xi_1(1D,5/2^+)\rangle=&
+0.143007|[[R_{00}^\rho R_{02}^\lambda]_{2}\chi_{1,\frac{1}{2}}]_{\frac{5}{2}}\rangle
+0.089769|[[R_{00}^\rho R_{02}^\lambda]_{2}\chi_{1,\frac{3}{2}}]_{\frac{5}{2}}\rangle
-0.186284|[[R_{01}^\rho R_{01}^\lambda]_{2}\chi_{0,\frac{1}{2}}]_{\frac{5}{2}}\rangle\\&
-0.907473|[[R_{02}^\rho R_{00}^\lambda]_{2}\chi_{1,\frac{1}{2}}]_{\frac{5}{2}}\rangle
-0.336574|[[R_{02}^\rho R_{00}^\lambda]_{2}\chi_{1,\frac{3}{2}}]_{\frac{5}{2}}\rangle.
\end{split}
\end{equation}
\end{flushleft}
\end{table*}

The configurations of $N=3$ states are very complex. They may be $1F$, radial excited $1P$, or their mixing. The calculated decay widths are presented in Table~\ref{tab:widthsN3}. In PDG~\cite{ParticleDataGroup:2024cfk}, there are two $N=3$ candidates, i.e., $\Omega(2380)$ and $\Omega(2470)$. For the $\Omega(2380)$, the measured mass is close to prediction of $\Omega|1/2^-\rangle_2$, as shown in Tables~\ref{tab:mass} and \ref{tab:widthsN3}. However, the $\Omega(2380)$ could be observed in $\Xi\bar{K}^0$ and occupy large relative branching ratio in $\Xi\bar{K}\pi$~\cite{Biagi:1985rn}, which may conflict the calculations in Table~\ref{tab:widthsN3}. In Table~\ref{tab:widthsN3}, the $\Omega|5/2^-\rangle_1$ is calculated with mass around 2.4~GeV, which is close to the mass of $\Omega(2380)$. On the other hand, the $\Omega|5/2^-\rangle_1$ has large $\Xi\bar{K}^*$ partial widths, which also match the measurement of $\Omega(2380)$. Thus the possible spin-parity of $\Omega(2380)$ is $5/2^-$.

Another $N=3$ candidate is $\Omega(2470)$. In PDG~\cite{ParticleDataGroup:2024cfk}, it should be a broad states but with large uncertainty, combine the calculations of masses and widths, it is possible the candidates of $\Omega|3/2^-\rangle_4$, $\Omega|7/2^-\rangle_1$, or $\Omega|9/2^-\rangle_1$.

Besides $\Omega(2380)$ and $\Omega(2470)$, Ref.~\cite{Aston:1987bb} reported a high excited state $\Omega(2561)$ in $\Xi^*\bar{K}$, which is not collected in PDG~\cite{ParticleDataGroup:2024cfk}. According to the central value of the experimental mass, $\Omega(2561)$ is possible $\Omega|3/2^-\rangle_6$ candidate. If we consider the errors of the measurement, the $\Omega|3/2^-\rangle_5$, $\Omega|5/2^-\rangle_4$, and $\Omega|7/2^-\rangle_2$ are also possible candidates.

In summary, the masses of $N=3$ are calculated in 2.38$\sim$2.57 GeV, with a large range. Besides the above candidates, there are many missing states. Most of these states are predicted not broad. We notice that some states have large partial $\Xi\bar{K}$ widths. And some states have considerable $\Xi^*\bar{K}$, $\Xi\bar{K}^*$, and $\Xi(1P)\bar{K}$ partial widths. Then the $\bar{K}^*$ can decay into $\bar{K}\pi$, the $\Xi^*$ and $\Xi(1P)$ could decay into $\Xi\pi$. In the secondary process, these states could be observed in the three-body channel $\Xi\bar{K}\pi$.

\section{Summary}\label{sec:Summary}

In this work, we conduct a systematic investigation of the properties of triple-strangeness $\Omega$ baryons. Employing a non-relativistic potential model within the framework of the Harmonic Oscillator Expansion Method, we calculate the mass spectra of these states. Subsequently, we utilize the QPC model to determine the partial and total decay widths. Our results provide critical insights into the assignments of experimentally observed $\Omega$ states and offer guidance for identifying currently missing states. The key findings are summarized as follows:

\begin{enumerate}
    \item The newly observed $\Omega(2109)$ is identified as a good $N=2$ candidate. Its mass and decay width suggest a plausible assignment as the $\Omega(2S,\,3/2^+)$ state. However, the measured mass lies approximately 50 MeV below our theoretical prediction. To reconcile this discrepancy, we propose two avenues: (i) Adjusting the confinement potential parameters to reduce the predicted mass; (ii) Conducting further searches for the $\Omega(2S,\,3/2^+)$ state near 2.16~GeV, closer to our calculated value.

    \item The $\Omega(2012)$ is a good $\Omega(1P,\,3/2^-)$ candidate. Since the mass of the $\Omega(2012)$ is close to the $\Xi^*\bar{K}$ threshold and three-body decay $\Xi\pi\bar{K}$ occupies large branching ratio, it is also a good hadronic molecular candidate. For higher excitations, the $\Omega(2250)$ is classified as an $N=2$ candidate, while the $\Omega(2380)$ and $\Omega(2470)$ are assigned to the $N=3$ tier.

    \item Beyond the observed states, our calculations predict numerous unobserved $N=2$ and $N=3$ $\Omega$ baryons. These states are expected to exhibit relatively narrow widths, making them accessible through channels such as $\Xi\bar{K}$, $\Xi^*\bar{K}$, $\Xi\bar{K}^*$, and $\Xi\bar{K}\pi$. We emphasize the importance of targeted searches in these final states.
\end{enumerate}

Recent observations by the ALICE Collaboration~\cite{ALICE:2025atb}, including the confirmation of $\Omega(2012)$, underscore the growing experimental capability to probe hyperon states. Our predictions for the masses, widths, and decay patterns of excited $\Omega$ baryons provide a road map for ongoing and future experiments at facilities such as Belle II, BESIII, ALICE, LHCb, and so on. The identification of missing states will not only refine our understanding of $\Omega$ spectroscopy but also test the robustness of quark potential models in describing high-mass baryonic systems.

\begin{acknowledgments}
This work is supported by the National Natural Science Foundation of China under Grant Nos. 12335001, 12247101, and 12405098,  the ‘111 Center’ under Grant No. B20063, the Natural Science Foundation of Gansu Province (No. 22JR5RA389, No. 25JRRA799), the Talent Scientific Fund of Lanzhou University, the fundamental Research Funds for the Central Universities, and the project for top-notch innovative talents of Gansu province.
\end{acknowledgments}

\section*{DATA AVAILABILITY}
The data that support the findings of this article are openly available~\cite{ParticleDataGroup:2024cfk,Belle:2018mqs,BESIII:2024eqk,Belle:2019zco,Belle:2022mrg}.

\end{document}